\documentstyle[12pt]{article}
\begin{document}

\immediate\write16{<WARNING: FEYNMAN macros work only with emTeX-dvivers 
                    (dviscr.exe, dvihplj.exe, dvidot.exe, etc.) >}
%% Macros for drawing Feynman graphs and other complex pictures
%% Designed by A.V.Voronin  (1993)
%% Steklov Math. Inst., e-mail: voronin@mph.mian.su 
%%
\newdimen\Lengthunit    
\newcount\Nhalfperiods  
%%%%%%%%%%%%%%%%%%%%%%%%%%%%%%%%%%%%%%%%%%%%%%%%%%%%%%%%%%%%%%%%%%
\Lengthunit = 1.5cm
\Nhalfperiods = 9
%%%%%%%%%%%%%%%%%%%%%%%%%%%%%%%%%%%%%%%%%%%%%%%%%%%%%%%%%%%%%%%%%%
\catcode`*=11
\newdimen\L*   \newdimen\d*   \newdimen\d**
\newdimen\dm*  \newdimen\dd*  \newdimen\dt*
\newdimen\a*   \newdimen\b*   \newdimen\c*
\newdimen\a**  \newdimen\b**
\newcount\k*   \newcount\l*   \newcount\m*
\newcount\n*   \newcount\dn*  \newcount\r*
\newcount\N*   \newcount\*one \newcount\*two  \*one=1 \*two=2
\newcount\*ths \*ths=1000   
%%%
\def\GRAPH(hsize=#1)#2{\hbox to #1\Lengthunit{#2\hfill}}
\def\Linewidth#1{\special{em:linewidth #1}}
\Linewidth{.4pt}
\def\sm*{\special{em:moveto}}
\def\sl*{\special{em:lineto}}
\newbox\spm*   \newbox\spl*
\setbox\spm*=\hbox{\sm*}
\setbox\spl*=\hbox{\sl*}
\def\mov#1(#2,#3)#4{\rlap{\L*=#1\Lengthunit\kern#2\L*\raise#3\L*\hbox{#4}}}
\def\mov*(#1,#2)#3{\rlap{\kern#1\raise#2\hbox{#3}}}
\def\lin#1(#2,#3){\rlap{\sm*\mov#1(#2,#3){\sl*}}}
\def\arr*(#1,#2,#3){\mov*(#1\dd*,#1\dt*){%
\sm*\mov*(#2\dd*,#2\dt*){\mov*(#3\dt*,-#3\dd*){\sl*}}%
\sm*\mov*(#2\dd*,#2\dt*){\mov*(-#3\dt*,#3\dd*){\sl*}}}}
\def\arrow#1(#2,#3){\rlap{\lin#1(#2,#3)\mov#1(#2,#3){%
\d**=-.012\Lengthunit\dd*=#2\d**\dt*=#3\d**%
\arr*(1,10,4)\arr*(3,8,4)\arr*(4.8,4.2,3)}}}
\def\arrlin#1(#2,#3){\rlap{\L*=#1\Lengthunit\L*=.5\L*%
\lin#1(#2,#3)\mov*(#2\L*,#3\L*){\arrow.1(#2,#3)}}}
\def\clap#1{\hbox to 0pt{\hss #1\hss}}
\def\ind(#1,#2)#3{\rlap{%
\d*=.1\Lengthunit\kern#1\d*\raise#2\d*\hbox{\lower2pt\clap{$#3$}}}}
\def\sh*(#1,#2)#3{\rlap{%
\dm*=\the\n*\d**\kern#1\dm*\raise#2\dm*\hbox{#3}}}
\def\calcnum*#1(#2,#3){\a*=1000sp\b*=1000sp\a*=#2\a*\b*=#3\b*%
\ifdim\a*<0pt\a*=-\a*\fi\ifdim\b*<0pt\b*=-\b*\fi%
\ifdim\a*>\b*\c*=.96\a*\advance\c*.4\b*%
\else\c*=.96\b*\advance\c*.4\a*\fi%
\k*=\a*\multiply\k*\k*\l*=\b*\multiply\l*\l*%
\m*=\k*\advance\m*\l*\n*=\c*\r*=\n*\multiply\n*\n*%
\dn*=\m*\advance\dn*-\n*\divide\dn*2\divide\dn*\r*%
\advance\r*\dn*%
\c*=\the\Nhalfperiods5sp\c*=#1\c*\ifdim\c*<0pt\c*=-\c*\fi%
\multiply\c*\r*\N*=\c*\divide\N*10000}
\def\dashlin#1(#2,#3){\rlap{\calcnum*#1(#2,#3)%
\d**=#1\Lengthunit\ifdim\d**<0pt\d**=-\d**\fi%
\divide\N*2\multiply\N*2\advance\N*1%
\divide\d**\N*\sm*\n*=\*one\sh*(#2,#3){\sl*}%
\loop\advance\n*\*one\sh*(#2,#3){\sm*}\advance\n*\*one\sh*(#2,#3){\sl*}%
\ifnum\n*<\N*\repeat}}
\def\dashdotlin#1(#2,#3){\rlap{\calcnum*#1(#2,#3)%
\d**=#1\Lengthunit\ifdim\d**<0pt\d**=-\d**\fi%
\divide\N*2\multiply\N*2\advance\N*1\multiply\N*2%
\divide\d**\N*\sm*\n*=\*two\sh*(#2,#3){\sl*}\loop%
\advance\n*\*one\sh*(#2,#3){\kern-1.48pt\lower.5pt\hbox{\rm.}}%
\advance\n*\*one\sh*(#2,#3){\sm*}%
\advance\n*\*two\sh*(#2,#3){\sl*}\ifnum\n*<\N*\repeat}}
\def\shl*(#1,#2)#3{\kern#1#3\lower#2#3\hbox{\sl*}}
\def\trianglin#1(#2,#3){\rlap{\toks0={#2}\toks1={#3}\calcnum*#1(#2,#3)%
\dd*=.57\Lengthunit\dd*=#1\dd*\divide\dd*\N*%
\d**=#1\Lengthunit\ifdim\d**<0pt\d**=-\d**\fi%
\multiply\N*2\divide\d**\N*\advance\N*-1\sm*\n*=\*one\loop%
\shl**{\dd*}\dd*=-\dd*\advance\n*2%
\ifnum\n*<\N*\repeat\n*=\N*\advance\n*1\shl**{0pt}}}
\def\wavelin#1(#2,#3){\rlap{\toks0={#2}\toks1={#3}\calcnum*#1(#2,#3)%
\dd*=.23\Lengthunit\dd*=#1\dd*\divide\dd*\N*%
\d**=#1\Lengthunit\ifdim\d**<0pt\d**=-\d**\fi%
\multiply\N*4\divide\d**\N*\sm*\n*=\*one\loop%
\shl**{\dd*}\dt*=1.3\dd*\advance\n*1%
\shl**{\dt*}\advance\n*\*one%
\shl**{\dd*}\advance\n*\*two%
\dd*=-\dd*\ifnum\n*<\N*\repeat\n*=\N*\shl**{0pt}}}
\def\w*lin(#1,#2){\rlap{\toks0={#1}\toks1={#2}\d**=\Lengthunit\dd*=-.12\d**%
\N*=8\divide\d**\N*\sm*\n*=\*one\loop%
\shl**{\dd*}\dt*=1.3\dd*\advance\n*\*one%
\shl**{\dt*}\advance\n*\*one%
\shl**{\dd*}\advance\n*\*one%
\shl**{0pt}\dd*=-\dd*\advance\n*1\ifnum\n*<\N*\repeat}}
\def\l*arc(#1,#2)[#3][#4]{\rlap{\toks0={#1}\toks1={#2}\d**=\Lengthunit%
\dd*=#3.037\d**\dd*=#4\dd*\dt*=#3.049\d**\dt*=#4\dt*\ifdim\d**>16mm%
\d**=.25\d**\n*=\*one\shl**{-\dd*}\n*=\*two\shl**{-\dt*}\n*=3\relax%
\shl**{-\dd*}\n*=4\relax\shl**{0pt}\else\ifdim\d**>4mm%
\d**=.5\d**\n*=\*one\shl**{-\dt*}\n*=\*two\shl**{0pt}%
\else\n*=\*one\shl**{0pt}\fi\fi}}
\def\d*arc(#1,#2)[#3][#4]{\rlap{\toks0={#1}\toks1={#2}\d**=\Lengthunit%
\dd*=#3.037\d**\dd*=#4\dd*\d**=.25\d**\sm*\n*=\*one\shl**{-\dd*}%
\n*=3\relax\sh*(#1,#2){\kern#2\dd*\lower#1\dd*\hbox{\sm*}}%
\n*=4\relax\shl**{0pt}}}
\def\arc#1[#2][#3]{\rlap{\Lengthunit=#1\Lengthunit%
\sm*\l*arc(#2.1914,#3.0381)[#2][#3]%
\mov(#2.1914,#3.0381){\l*arc(#2.1622,#3.1084)[#2][#3]}%
\mov(#2.3536,#3.1465){\l*arc(#2.1084,#3.1622)[#2][#3]}%
\mov(#2.4619,#3.3086){\l*arc(#2.0381,#3.1914)[#2][#3]}}}
\def\dasharc#1[#2][#3]{\rlap{\Lengthunit=#1\Lengthunit%
\d*arc(#2.1914,#3.0381)[#2][#3]%
\mov(#2.1914,#3.0381){\d*arc(#2.1622,#3.1084)[#2][#3]}%
\mov(#2.3536,#3.1465){\d*arc(#2.1084,#3.1622)[#2][#3]}%
\mov(#2.4619,#3.3086){\d*arc(#2.0381,#3.1914)[#2][#3]}}}
\def\wavearc#1[#2][#3]{\rlap{\Lengthunit=#1\Lengthunit%
\w*lin(#2.1914,#3.0381)%
\mov(#2.1914,#3.0381){\w*lin(#2.1622,#3.1084)}%
\mov(#2.3536,#3.1465){\w*lin(#2.1084,#3.1622)}%
\mov(#2.4619,#3.3086){\w*lin(#2.0381,#3.1914)}}}
\def\shl**#1{\c*=\the\n*\d**\d*=#1%
\a*=\the\toks0\c*\b*=\the\toks1\d*\advance\a*-\b*%
\b*=\the\toks1\c*\d*=\the\toks0\d*\advance\b*\d*%
\rlap{\kern\a*\raise\b*\copy\spl*}}
\def\wlin*#1(#2,#3)[#4]{\rlap{\toks0={#2}\toks1={#3}%
\c*=#1\l*=\c*\c*=.01\Lengthunit\m*=\c*\divide\l*\m*%
\c*=\the\Nhalfperiods5sp\multiply\c*\l*\N*=\c*\divide\N*\*ths%
\divide\N*2\multiply\N*2\advance\N*1%
\dd*=.002\Lengthunit\dd*=#4\dd*\multiply\dd*\l*\divide\dd*\N*%
\d**=#1\multiply\N*4\divide\d**\N*\sm*\n*=\*one\loop%
\shl**{\dd*}\dt*=1.3\dd*\advance\n*\*one%
\shl**{\dt*}\advance\n*\*one%
\shl**{\dd*}\advance\n*\*two%
\dd*=-\dd*\ifnum\n*<\N*\repeat\n*=\N*\shl**{0pt}}}
\def\wavebox#1{\setbox0\hbox{#1}%
\a*=\wd0\advance\a*14pt\b*=\ht0\advance\b*\dp0\advance\b*14pt%
\hbox{\kern9pt%
\mov*(0pt,\ht0){\mov*(-7pt,7pt){\wlin*\a*(1,0)[+]\wlin*\b*(0,-1)[-]}}%
\mov*(\wd0,-\dp0){\mov*(7pt,-7pt){\wlin*\a*(-1,0)[+]\wlin*\b*(0,1)[-]}}%
\box0\kern9pt}}
\def\rectangle#1(#2,#3){%
\lin#1(#2,0)\lin#1(0,#3)\mov#1(0,#3){\lin#1(#2,0)}\mov#1(#2,0){\lin#1(0,#3)}}
\def\dashrectangle#1(#2,#3){\dashlin#1(#2,0)\dashlin#1(0,#3)%
\mov#1(0,#3){\dashlin#1(#2,0)}\mov#1(#2,0){\dashlin#1(0,#3)}}
\def\waverectangle#1(#2,#3){\L*=#1\Lengthunit\a*=#2\L*\b*=#3\L*%
\ifdim\a*<0pt\a*=-\a*\def\x*{-1}\else\def\x*{1}\fi%
\ifdim\b*<0pt\b*=-\b*\def\y*{-1}\else\def\y*{1}\fi%
\wlin*\a*(\x*,0)[-]\wlin*\b*(0,\y*)[+]%
\mov#1(0,#3){\wlin*\a*(\x*,0)[+]}\mov#1(#2,0){\wlin*\b*(0,\y*)[-]}}
\def\calcparab*{%
\ifnum\n*>\m*\k*=\N*\advance\k*-\n*\else\k*=\n*\fi%
\a*=\the\k* sp\a*=10\a*\b*=\dm*\advance\b*-\a*\k*=\b*%
\a*=\the\*ths\b*\divide\a*\l*\multiply\a*\k*%
\divide\a*\l*\k*=\*ths\r*=\a*\advance\k*-\r*%
\dt*=\the\k*\L*}
\def\arcto#1(#2,#3)[#4]{\rlap{\toks0={#2}\toks1={#3}\calcnum*#1(#2,#3)%
\dm*=135sp\dm*=#1\dm*\d**=#1\Lengthunit\ifdim\dm*<0pt\dm*=-\dm*\fi%
\multiply\dm*\r*\a*=.3\dm*\a*=#4\a*\ifdim\a*<0pt\a*=-\a*\fi%
\advance\dm*\a*\N*=\dm*\divide\N*10000%
\divide\N*2\multiply\N*2\advance\N*1%
\L*=-.25\d**\L*=#4\L*\divide\d**\N*\divide\L*\*ths%
\m*=\N*\divide\m*2\dm*=\the\m*5sp\l*=\dm*%
\sm*\n*=\*one\loop\calcparab*\shl**{-\dt*}%
\advance\n*1\ifnum\n*<\N*\repeat}}
\def\arrarcto#1(#2,#3)[#4]{\L*=#1\Lengthunit\L*=.54\L*%
\arcto#1(#2,#3)[#4]\mov*(#2\L*,#3\L*){\d*=.457\L*\d*=#4\d*\d**=-\d*%
\mov*(#3\d**,#2\d*){\arrow.02(#2,#3)}}}
\def\dasharcto#1(#2,#3)[#4]{\rlap{\toks0={#2}\toks1={#3}\calcnum*#1(#2,#3)%
\dm*=\the\N*5sp\a*=.3\dm*\a*=#4\a*\ifdim\a*<0pt\a*=-\a*\fi%
\advance\dm*\a*\N*=\dm*%
\divide\N*20\multiply\N*2\advance\N*1\d**=#1\Lengthunit%
\L*=-.25\d**\L*=#4\L*\divide\d**\N*\divide\L*\*ths%
\m*=\N*\divide\m*2\dm*=\the\m*5sp\l*=\dm*%
\sm*\n*=\*one\loop%
\calcparab*\shl**{-\dt*}\advance\n*1%
\ifnum\n*>\N*\else\calcparab*%
\sh*(#2,#3){\kern#3\dt*\lower#2\dt*\hbox{\sm*}}\fi%
\advance\n*1\ifnum\n*<\N*\repeat}}
\def\*shl*#1{%
\c*=\the\n*\d**\advance\c*#1\a**\d*=\dt*\advance\d*#1\b**%
\a*=\the\toks0\c*\b*=\the\toks1\d*\advance\a*-\b*%
\b*=\the\toks1\c*\d*=\the\toks0\d*\advance\b*\d*%
\rlap{\kern\a*\raise\b*\copy\spl*}}
\def\calcnormal*#1{%
\b**=10000sp\a**=\b**\k*=\n*\advance\k*-\m*%
\multiply\a**\k*\divide\a**\m*\a**=#1\a**\ifdim\a**<0pt\a**=-\a**\fi%
\ifdim\a**>\b**\d*=.96\a**\advance\d*.4\b**%
\else\d*=.96\b**\advance\d*.4\a**\fi%
\d*=.01\d*\r*=\d*\divide\a**\r*\divide\b**\r*%
\ifnum\k*<0\a**=-\a**\fi\d*=#1\d*\ifdim\d*<0pt\b**=-\b**\fi%
\k*=\a**\a**=\the\k*\dd*\k*=\b**\b**=\the\k*\dd*}
\def\wavearcto#1(#2,#3)[#4]{\rlap{\toks0={#2}\toks1={#3}\calcnum*#1(#2,#3)%
\c*=\the\N*5sp\a*=.4\c*\a*=#4\a*\ifdim\a*<0pt\a*=-\a*\fi%
\advance\c*\a*\N*=\c*\divide\N*20\multiply\N*2\advance\N*-1\multiply\N*4%
\d**=#1\Lengthunit\dd*=.012\d**\ifdim\d**<0pt\d**=-\d**\fi\L*=.25\d**%
\divide\d**\N*\divide\dd*\N*\L*=#4\L*\divide\L*\*ths%
\m*=\N*\divide\m*2\dm*=\the\m*0sp\l*=\dm*%
\sm*\n*=\*one\loop\calcnormal*{#4}\calcparab*%
\*shl*{1}\advance\n*\*one\calcparab*%
\*shl*{1.3}\advance\n*\*one\calcparab*%
\*shl*{1}\advance\n*2%
\dd*=-\dd*\ifnum\n*<\N*\repeat\n*=\N*\shl**{0pt}}}
\def\triangarcto#1(#2,#3)[#4]{\rlap{\toks0={#2}\toks1={#3}\calcnum*#1(#2,#3)%
\c*=\the\N*5sp\a*=.4\c*\a*=#4\a*\ifdim\a*<0pt\a*=-\a*\fi%
\advance\c*\a*\N*=\c*\divide\N*20\multiply\N*2\advance\N*-1\multiply\N*2%
\d**=#1\Lengthunit\dd*=.012\d**\ifdim\d**<0pt\d**=-\d**\fi\L*=.25\d**%
\divide\d**\N*\divide\dd*\N*\L*=#4\L*\divide\L*\*ths%
\m*=\N*\divide\m*2\dm*=\the\m*0sp\l*=\dm*%
\sm*\n*=\*one\loop\calcnormal*{#4}\calcparab*%
\*shl*{1}\advance\n*2%
\dd*=-\dd*\ifnum\n*<\N*\repeat\n*=\N*\shl**{0pt}}}
\edef\hr*#1{\clap{\vrule width#1\Lengthunit height.1pt}}
\def\shade#1[#2]{\rlap{\Lengthunit=#1\Lengthunit%
\mov(0,#2.05){\hr*{.994}}\mov(0,#2.1){\hr*{.980}}%
\mov(0,#2.15){\hr*{.953}}\mov(0,#2.2){\hr*{.916}}%
\mov(0,#2.25){\hr*{.867}}\mov(0,#2.3){\hr*{.798}}%
\mov(0,#2.35){\hr*{.715}}\mov(0,#2.4){\hr*{.603}}%
\mov(0,#2.45){\hr*{.435}}}}
\def\dshade#1[#2]{\rlap{%
\Lengthunit=#1\Lengthunit\if#2-\def\t*{+}\else\def\t*{-}\fi%
\mov(0,\t*.025){%
\mov(0,#2.05){\hr*{.995}}\mov(0,#2.1){\hr*{.988}}%
\mov(0,#2.15){\hr*{.969}}\mov(0,#2.2){\hr*{.937}}%
\mov(0,#2.25){\hr*{.893}}\mov(0,#2.3){\hr*{.836}}%
\mov(0,#2.35){\hr*{.760}}\mov(0,#2.4){\hr*{.662}}%
\mov(0,#2.45){\hr*{.531}}\mov(0,#2.5){\hr*{.320}}}}}
\def\vdot{\rlap{\kern-1.9pt\lower1.8pt\hbox{$\scriptstyle\bullet$}}}
\def\vtimes{\rlap{\kern-3pt\lower1.8pt\hbox{$\scriptstyle\times$}}}
\def\vDot{\rlap{\kern-2.3pt\lower2.7pt\hbox{$\bullet$}}}
\def\vTimes{\rlap{\kern-3.6pt\lower2.4pt\hbox{$\times$}}}
\catcode`*=12
\newcount\CatcodeOfAtSign
\CatcodeOfAtSign=\the\catcode`\@
\catcode`\@=11
\newcount\n@ast
\def\n@ast@#1{\n@ast=0\relax\get@ast@#1\end}
\def\get@ast@#1{\ifx#1\end\let\next\relax\else%
\ifx#1*\advance\n@ast1\fi\let\next\get@ast@\fi\next}
\newif\if@up \newif\if@dwn
\def\up@down@#1{\@upfalse\@dwnfalse%
\if#1u\@uptrue\fi\if#1U\@uptrue\fi\if#1+\@uptrue\fi%
\if#1d\@dwntrue\fi\if#1D\@dwntrue\fi\if#1-\@dwntrue\fi}
\def\halfcirc#1(#2)[#3]{{\Lengthunit=#2\Lengthunit\up@down@{#3}%
\if@up\mov(0,.5){\arc[-][-]\arc[+][-]}\fi%
\if@dwn\mov(0,-.5){\arc[-][+]\arc[+][+]}\fi%
\def\lft{\mov(0,.5){\arc[-][-]}\mov(0,-.5){\arc[-][+]}}%
\def\rght{\mov(0,.5){\arc[+][-]}\mov(0,-.5){\arc[+][+]}}%
\if#3l\lft\fi\if#3L\lft\fi\if#3r\rght\fi\if#3R\rght\fi%
\n@ast@{#1}%
\ifnum\n@ast>0\if@up\shade[+]\fi\if@dwn\shade[-]\fi\fi%
\ifnum\n@ast>1\if@up\dshade[+]\fi\if@dwn\dshade[-]\fi\fi}}
\def\halfdashcirc(#1)[#2]{{\Lengthunit=#1\Lengthunit\up@down@{#2}%
\if@up\mov(0,.5){\dasharc[-][-]\dasharc[+][-]}\fi%
\if@dwn\mov(0,-.5){\dasharc[-][+]\dasharc[+][+]}\fi%
\def\lft{\mov(0,.5){\dasharc[-][-]}\mov(0,-.5){\dasharc[-][+]}}%
\def\rght{\mov(0,.5){\dasharc[+][-]}\mov(0,-.5){\dasharc[+][+]}}%
\if#2l\lft\fi\if#2L\lft\fi\if#2r\rght\fi\if#2R\rght\fi}}
\def\halfwavecirc(#1)[#2]{{\Lengthunit=#1\Lengthunit\up@down@{#2}%
\if@up\mov(0,.5){\wavearc[-][-]\wavearc[+][-]}\fi%
\if@dwn\mov(0,-.5){\wavearc[-][+]\wavearc[+][+]}\fi%
\def\lft{\mov(0,.5){\wavearc[-][-]}\mov(0,-.5){\wavearc[-][+]}}%
\def\rght{\mov(0,.5){\wavearc[+][-]}\mov(0,-.5){\wavearc[+][+]}}%
\if#2l\lft\fi\if#2L\lft\fi\if#2r\rght\fi\if#2R\rght\fi}}
\def\Circle#1(#2){\halfcirc#1(#2)[u]\halfcirc#1(#2)[d]\n@ast@{#1}%
\ifnum\n@ast>0\clap{\vrule width#2\Lengthunit height.1pt}\fi}
\def\wavecirc(#1){\halfwavecirc(#1)[u]\halfwavecirc(#1)[d]}
\def\dashcirc(#1){\halfdashcirc(#1)[u]\halfdashcirc(#1)[d]}
\catcode`\@=\the\CatcodeOfAtSign


%%%%%%%%%%%%%%%%%%%%%%%%%%%%%%%%%%%%%%%%%%%%%%%%%%%%%%%%%%%%%%%%%%%%%%%%%%%%%%%%%%%%%%
\def\be{\begin{equation}}
\def\ee{\end{equation}}
\def\bea{\begin{eqnarray}}
\def\eea{\end{eqnarray}}
\def\fmn{{\cal F}_{\mu\nu}}
\def\am{{\cal A}_{\mu}}
\def\ama{\am^{a}}
\def\qm{q_{\mu}}
\def\qma{\qm^{a}}
\def\sn{S_{n}^{\Lambda}}
\def\kongo{\frac{1}{2}\int\!\int\frac{\delta^{2}\sn}{\delta\ama(x)\delta{\cal
A}_{\nu}^{b}(y)}\qma(x) q_{\nu}^{b}(y)dxdy}
\def\kon{\Re_{n} (\am,\qm)}
\def\gft{\frac{1}{4\alpha}{\bf tr}\{f_{n}(\frac{\Box}{\Lambda^2})\,\partial_{\mu}\am \}^{2}}
\def\gftq{\frac{1}{4\alpha}{\bf tr}\{f_n (\frac{\Box}{\Lambda^2})\,\partial_{\mu}\qm \}^2}
\def\za{{\cal Z}_{\alpha}}
\def\zo{{\cal Z}_{0}}
\def\zinv{{\cal Z}_{\mbox{inv}}}
\def\gda{\det{\cal M}({\cal A})}
\def\dd{\prod_{x}} \def\pdm{\partial_{\mu}} \def\nm{\nabla_{\mu}}
\def\Fn{F_{n}\Bigl(\frac{\nabla^2}{\Lambda^2}\Bigr)}
\def\bFn{\frac{1}{2\beta} \int [\Fn \nm \qm]^{2}dx}
\def\cj{c_{\jmath}}
\def\dnm{det (\nabla^2-M^2)}
\def\dnmj{det^{\cj} (\nabla^2 - M_{\jmath}^2)}
\def\dqm{det_{n}^{-1/2}Q_{\beta}(\am;M^2;F_{n})}
\def\dqmj{det_{n}^{-\cj /2}Q_{\beta}(\am;M_{\jmath}^{2};F_{n})}
\def\zlm{{\cal Z}^{n}_{\Lambda,M^2}[J]}
\def\hact{\Bigl[\frac{\delta S_0}{\delta {\cal A}_{\rho}(z)}+\nabla_{\rho}
\Box^2\partial {\cal A}(z)\Bigr]}
\def\soa{\frac{\delta^2S_0}{\delta \am (x)\delta {\cal A}_{\rho}(z)}}
\def\opdet{\Bigl[\soa+\nabla_{\rho}\Box^2\pdm \delta(x-z)\Bigr]}
\def\detdet{det^{1/2}\Bigl[\Box^2\nabla^2\Box^2+\alpha^{-1}\Lambda^{10}\Bigr]}
\def\finv{ F\Bigl(\frac{\nabla^2}{\Lambda^2}\Bigr)}
\def\indet{\frac{1}{\Lambda^5}\soa+\frac{\nm}{\Lambda}\finv\nabla_\rho\delta
(x-z)}
\def\klio{det^{1/2}\Bigl[\nabla^2+\alpha^{-1}\Box^{-4}\Lambda^{10}\Bigr]}
\def\cov{\frac{1}{\Lambda}(\nm\nabla_\rho-\nabla^2g_{\mu\rho})\delta(x-z)}

%%%%%%%%%%%%%%%%%%%%%%%%%%%%%%%%%%%%%%%%%%%%%%%%%%%%%%%%%%%%%%%%%%%%%%%%%%%%%%%%%

\title{
\bf \mbox{} \\ HIGHER COVARIANT DERIVATIVE REGULARIZATION
REVISITED.} \vspace{.5cm} \author{{\bf T.D.Bakeyev \ and \ A.A.Slavnov
\thanks{E-mail: slavnov@class.mian.su} }
\vspace{.5cm} \date{} \\ {\it Steclov Mathematical Institute, Russian
Academy of Sciences,} \\ {\it Vavilov st.42, GSP-1, 117966 Moscow, RUSSIA}
\\ {\it and} \\ {\it Moscow State University.}}

\maketitle 
\vspace{-10cm}
{\hfill SMI-02-96}
\vspace{8cm}

\vspace{2cm}
 \begin{abstract} The method of higher covariant
derivative regularization of gauge theories is reviewed. The objections
raised in the literature last years are discussed and the consistency of
the method is proven. New approach to regularization of overlapping divergencies
is developped.
\end{abstract}
\vspace{2cm}

 \section{Introduction.}  In this paper we
review the method of higher covariant derivative regularization of gauge
theories \cite{Sl1,Zinn} supplemented by the additional Pauli-Villars (PV)
type regularization as proposed in \cite{Sl3} (see also \cite{SlF}).  We
analyze the objections raised in papers \cite{Warr,Sen,Ruiz} and show that
although indeed some minor modifications of the original scheme are
needed, the general method is self-consistent, provides the
gauge invariant regularization to all orders in perturbation theory and on
the other hand may serve as a starting point for nonperturbative
calculations.

 The problem of invariant regularization is of extreme
importance both from the point of view of practical calculations and for
the general study of symmetry properties of renormalized quantum theory.  It
is widely believed that for anomaly free models an invariant
regularization do exist although no general theorem was proven.

The method mostly used so far for calculations in gauge
invariant models was the dimensional regularization \cite{Dim}. However
dimensional regularization is not applicable to chiral and supersymmetric models which
are very important from the point of view of applications. Moreover there
is no obvious generalization of this method to nonperturbative
calculations, in particular dealing with topological aspects of a theory,
as the dimensional regularization is formulated in terms of perturbative
 Feynmann diagrams.

The most natural nonperturbative regularization is
provided probably by the lattice formulation. But this approach also meets
some difficulties in treating topological and chiral models. It is also
not very practical for weak coupling calculations due to appearance of new
vertices and lack of Lorentz (rotational) invariance.

An alternative regularization scheme which may be implemented as a modification of
classical Lagrangian and therefore has a nonperturbative meaning is the
higher covariant derivative method. This method also has an advantage of
being applicable to chiral and supersymmetric models, but due to a rather
complicated structure of regularized Lagrangian it was used mainly for
general proofs and not in practical calculations. However it seems that
nowadays with the need in precision calculations of electroweak processes on
one side and the big progress in computer facilities on the other side,
this method may become a real practical tool. For that reason some tests of
the procedure were carried out last years, which raised some controversy
in the literature.

So we feel it is worthwhile to review the method and to discuss the
problems raised in the process of it's testing.

\section{General idea of the method.}
The most simple way to regularize the theory is to modify the propagators
by introducing into Lagrangian higher derivative terms. However this procedure
breaks gauge invariance. To preserve the symmetry it was proposed to
modify the Yang-Mills(YM) Lagrangian by adding the terms containing
higher covariant derivatives \cite{Sl1,Zinn}, e.g.

\be
{\cal L}_{YM} \rightarrow {\cal L}_{n}^{\Lambda}=\frac{1}{8}
{\bf tr}\{\fmn^2+\frac {1} {\Lambda^{2n}}\bigl(\nabla^n\fmn\bigr)^{2}\}
\label{1}\ee

Here $\fmn$ is the usual curvature tensor and $\nabla$ is the
covariant derivative:

\be \nabla_{\alpha} \fmn=\partial_{\alpha}\fmn+\bigl[{\cal A}_{\alpha},\fmn\bigr]
\label{2} \ee

This regularization improves the ultraviolet behaviour of the YM field
propagator, provided the gauge fixing term is chosen in the form:

\be\gft \label{3}\ee
where $f_{n}$ is a polinomial of order $\geq \frac{n}{2}$.

Using the Lagrangian (\ref{1}) with the gauge fixing term (\ref{3}) one
easily sees that the divergency index of arbitrary diagram is equal to:

\be \omega_{n}=4-2n(I-1)-E_{\cal A}-(n+1)E_{C} \label{4}\ee
where $I$ is the number of loops, $E_{\cal A}$ and $E_{C}$ are the numbers
of external gauge and ghost field lines correspondingly. Therefore for
$2n\geq 4$ we got the theory with a finite number of divergent diagrams.
Namely, the only divergent graphs are the one loop diagrams with $E_{\cal
A}=2,3,4$ and $E_{C}=0$.

This procedure makes convergent all multiloop diagrams in arbitrary gauge
theory, however the one loop diagrams require some additional regularization.

It was proposed in the paper \cite{Sl3} (see also \cite{SlF}) that such
a regularization may be provided by a modified PV procedure.

The key observation was the following. In a higher covariant derivative
gauge theory the remaining divergency must have a manifestly gauge invariant
structure. The corresponding counterterm is:

\be Z tr\{\fmn\fmn\}\label{5}\ee

It follows directly from Slavnov-Taylor identities \cite{Sl5,Tail} and the
fact that the ghost fields and vertex renormalizations in a higher
covariant gauge theory are finite. It suggests that these divergencies may be
regularized by adding some gauge invariant interaction providing analogous counterterms
with the opposite sign.

The formal scheme looks as follows. The total contribution of one loop diagrams
with external gauge field lines may be represented by:

\be \za[\am]=\int\exp \Bigl \{\imath[\kon+\int\gftq dx ]\Bigr \}\gda\dd
D\qm \label{6}\ee
  where $\kon=\kongo$, $\sn$ is the regularized action corresponding to
  the Lagrangian (\ref{1}).

This functional is not invariant with respect to the gauge transformations
of the fields $\am$, but it's divergent part is gauge invariant. We
firstly demonstrate it for the special case of the Lorentz gauge
$\alpha=0$ and then consider the general case.

The functional $\zo$ given by the equation:

\be  \zo[{\cal A}]=\int\hbox{e}^{\imath \kon}\gda\dd\delta(\pdm\qm) D \qm
\label{7}\ee
may be transformed to the following form:

\be \ln\zo=\ln\zinv+[\mbox{finite part}]\label{8}\ee
where $\zinv[\am]$ is a manifestly gauge invariant functional. It can be
done by passing in the eq.(\ref{7}) to the covariant background gauge. To
perform the transition we multiply eq.(\ref{7}) by ''unity'':

\be det \Fn det\nabla^2\int\delta \Bigl(\Fn\nm(\qm+\nm u)-W(x)\Bigr)\dd
Du(x)=1 \label{9} \ee where $\Fn$ is a polynomial of degree:
\be\frac{n+1}{4}<\,deg F_{n}\,\leq \frac{n}{2}\label{10}\ee
\[\lim_{\Lambda\rightarrow\infty}\Fn=1 \]
and $\nm$ denotes a covariant derivative with respect to the field $\am$.
Changing variables:

\be\qm\rightarrow\qm\,-\,\nm u \label{11}\ee
we have:

\bea
&&\zo =\int \exp\Bigl\{\imath \Re_{n}(\am;\qm-\nm u)\Bigr\}\gda
det\nabla^{2}det\Fn \nonumber \\ &&\dd \delta (\pdm(\qm-\nm u))\delta(\Fn
\nabla_{\nu} q_{\nu} - W(x))DuD\qm \label{12} \eea

The functional $\zo$ does not depend on $W$, so we can integrate it over $W$
with the weight $\exp\{\frac{\imath}{2\beta}\int W^2 (x)dx\}$. Integrating
over $W$ and $u$ we get:

\bea
&&\zo=\int \exp\Bigl \{\imath [\Re_{n}(\am;\qm -\nm {\cal M}({\cal A})^{-1}
\partial_{\rho} q_{\rho})+\bFn ]\Bigr\} \nonumber \\ &&det \nabla^2 det \Fn
\dd D\qm(x) \label{13} \eea

The functional:

\be \zinv [{\cal A}]=\int \hbox{e}^{\imath [\kon +\bFn]}det \nabla^2 det \Fn
\dd D\qm(x) \label{14} \ee is invariant with respect to the gauge
transformations of fields $\am$, as the exponent does not change under
simultaneous transformations:

\be \left\{ \begin{array}{ll}
             \am\rightarrow\am +[\am,\epsilon] +\pdm \epsilon \\
             \qm\rightarrow\qm +[\qm,\epsilon]
             \end{array}
    \right. \label{15} \ee

At the same time it's connected part differs from (\ref{13}) only by
finite terms. Indeed, since $\sn$ is gauge invariant, we have:

\be \int \frac{\delta^2 \sn}{\delta \ama(x) \delta {\cal A}_{\nu}^{b}(y)}
 [\nm \phi (x)]^{a} dx =gt^{abc} \frac{\delta \sn}{\delta
{\cal A}_{\nu}^{a}(y)}\phi^{c}(y) \label{16} \ee

The additional terms in the exponent of eq. (\ref{13}) which are not
present in (\ref{14}) can be rewritten as follows:

\be gt^{abc} \int \frac{\delta \sn}{\delta\ama (x)}[\qm -\frac{1}{2}\nm
{\cal M}^{-1} \partial_{\rho} q_{\rho}]^{b}_{x} [{\cal M}^{-1}
\partial_{\sigma} q_{\sigma}]^{c}_{x} \label{17} \ee Due to the fact that
some derivatives in the eq.(\ref{17}) act on the external fields $\am$,
and the maximal number of derivatives acting on $\qm$ is $(n+1)$, the
corresponding diagrams are not divergent if the condition (\ref{10})
holds.

In the same way one can prove that the connected part of $\za$ differs
from $\zo$ by finite terms. Multiplying $\zo$ by "unity":

\be \gda \int \dd \delta \Bigl(\pdm(\qm +\nm u) - W(x)\Bigr) Du(x)=1
\label{18} \ee and integrating it over $W$ with the weight $\exp
\{\frac{\imath}{2\alpha} \int [f_{n}(\frac{\Box}{\Lambda^{2}})W(x)]^{2} dx
\}$ one can prove it in complete analogy with the discussion above.

The transformation of $\za$ described above makes the gauge invariance of
it's divergent part manifest. It also suggests a natural gauge invariant
regularization of this functional. Under gauge transformations (\ref{15})
the fields $\qm$ transform homogeneously. Therefore without breaking the
gauge invariance one can add mass terms for the fields $\qm$. It allows to
use for regularization of the functional (\ref{14}) the PV procedure. The
regularization looks as follows:

\be \zinv [\am] \rightarrow \zinv [\am] det^{-1} \Fn \prod_{\jmath}[
\dnmj \times \dqmj] \label{19} \ee
where:

\bea
\dqm &=& \int\exp\Bigl\{\imath [\kon+\bFn - \nonumber \\
&&-\frac{M^2}{2} \int (\qma)^2 dx]\Bigr\}  \dd D\qm(x)
 \label{20} \eea
and PV conditions hold:

\be 1+\sum_{\jmath} \cj =0 \label{21} \ee
\be \sum_{\jmath} \cj M_{\jmath}^2=0 \label{21a} \ee
(In higher derivative theory the condition (\ref{21}) is sufficient for
the regularization of $det_{n}^{-1/2}Q_{\beta}$ as a single subtraction like
\be \frac{1}{k^2+\Lambda^{-2n}k^{2n+2}} -
\frac{1}{k^2+\Lambda^{-2n}k^{2n+2}+M^2} \ee
makes the integral convergent if $n \geq1$.)
Perturbative expansion of the expression (\ref{19}) generates together
with the loops formed by the original fields $\qm$ analogous loops of
PV fields.  Due to eq.(\ref{21},\ref{21a}) leading ultraviolet asymptotics
cancel and the corresponding integrals are convergent. Regularizing terms
are obviously gauge invariant.

To remove the regularization one should take the limit $\Lambda \rightarrow
\infty$,$M\rightarrow\infty$. In this limit the PV fields
decouple from the physical fields and contribute only to local
counterterms. This is true for any $\beta \neq 0$.  However the case
$\beta=0$ is singular and needs more carefull study. This fact was
overlooked in papers \cite{Sl3,SlF} and if one applies directly the eqs.
written in refs. \cite{Sl3,SlF} for the Lorentz gauge $\beta=0$ to
calculations of one loop results one gets a wrong result. It was
demonstrated explicitely in ref. \cite{Ruiz} which lead the authors to the
claim that the higher covariant derivative method is inconsistent. As was
pointed out by M. Asorey and M. Falceto \cite{AsF} the discrepancy does
not mean inconsistency of the method but is due to the singular character
of the Lorentz gauge. In this gauge in the limit $M\rightarrow\infty$
the regularizing fields do not decouple completely. It is most easily seen
by rescaling the fields $\qm$ in the eq.  (\ref{20}): $\qm \rightarrow
\frac{1}{M} \qm$. For $\beta \neq 0$ after this rescaling all the terms
exept for $\qm^{2}$ vanish in the limit $M \rightarrow 0$ and the
integral over $\qm$ gives nonessential constant.
However for $\beta=0$ this rescaling does not kill the gauge fixing term
and the integration over $\qm$ produces additional factor which survives
in the limit $M \rightarrow \infty$. As was shown in ref. \cite{AsF} to
get the correct result one has to subtract this factor. The analog of eqs.
(\ref{19},\ref{20}) for the case of covariant Lorentz gauge looks as follows:

\be \zinv \rightarrow  \zinv det^{-1/2}\nabla^2
\prod_{\jmath}[det_{n}^{\;\,-\cj/2}Q_{\beta=0}(\am;M_{\jmath}^2)\times
det^{\cj/2}(\nabla^2-M_{\jmath}^2)] \label{22} \ee
where:

\be det_{n}^{-1/2}Q_{\beta=0}(\am;M^2)=\int \exp\Bigl\{\imath
[\kon -\frac{M^2}{2}\int(\qma)^2 dx]\Bigr\} \dd \delta (\nm \qm)D
\qm(x) \label{23} \ee
As in the case $\beta \neq 0$ the regularizing terms are obviously gauge
invariant.

Taking into account that as was proven above:

\be \ln \za=\ln \zinv +[\mbox{finite part}] \label{24} \ee
one would expect that a natural gauge invariant regularization of $\za$ may
be chosen in the form:

\bea &&\zlm =\int \exp\Bigl\{\imath \int
[{\cal L}_{n}^{\Lambda} + J_{\mu}^{a}\ama+ \gft]dx\Bigr\}\gda \nonumber \\
&&\cdot det^{-1} \Fn \prod_{\jmath}[\dnmj \dqmj] \dd D \am(x)
\label{25}\eea

However the straightforward application of this equation is ambigous.
The point is that individual diagrams generated
by the perturbative expansion of the functional (\ref{25}) still may be
divergent.  The formal derivation given above refers to the sum of all
diagrams of a given order in perturbation theory with fixed numbers of
external lines. The statement that $\ln \za$ differs of $\ln \zinv$ by
finite terms is true for the sum of the diagrams of a given order and not
for individual diagrams.  To give a precise meaning to the expression
(\ref{25}) we have to specify the procedure of summation of divergent
diagrams. In fact the same problem exists for the usual PV regularization
and in this case it is solved in the following way.

Let us denote the propagator of particles with the mass $M$ by $D_{M^2}^{ab}(p^2)$
and the vertex factor by $\Gamma^{ab}$. PV regularized expression for a
loop with $n$ vertices looks as follows:

\bea &&\int dp [D_{M^2}^{a_{1} b_{1}}(p)\Gamma^{b_{1} a_{2}}\cdot \ldots
\cdot D_{M^2}^{a_{n} b_{n}}(p+k_{n-1}) \Gamma^{b_{n} a_{1}} +\nonumber \\
&&+\sum_{\jmath=1}^{J} \cj [D_{M^{2}_{\jmath}}^{a_{1}
b_{1}}(p)\Gamma^{b_{1} a_{2}}\cdot \ldots \cdot
D_{M_{\jmath}^{2}}^{a_{n} b_{n}}(p+k_{n-1})\Gamma^{b_{n} a_{1}}]
\label{26} \eea

It corresponds to the sum of similar loop diagrams describing the propagation
of particles with masses $M,M_1,\ldots,M_J$. Two rules are assumed:
a) all internal momenta have to be arranged in the same way in each diagram;
b)the sum is to be taken before integration.

Unfortunately this simple recipe does not work for the eq.(\ref{25}). The
diagrams generated by ${\cal L}_{n}^\Lambda$ and by the PV fields have a
different structure.  To make sence of this expression one has to
introduce some preregularization (PR) procedure which makes the individual
diagrams finite and the summation unambigous. In ref.\cite{AsF2} the momentum
cutoff procedure for all internal lines was used as a PR. The necessity of PR in
higher covariant derivative PV regularization was also emphasized in
ref.\cite{Ruiz}.

After the summation is performed the PR has to be removed.  Of course one
must show that this procedure does not break gauge invariance and after
the PR is removed the functional $\zlm$ satisfies Slavnov-Taylor identities.

The second objection is
related to the problem of diagrams with divergent subgraphs, in particular
with overlapping divergencies. Although the one loop diagrams with
external $\am$ lines generated by the
eq. (\ref{25}) are finite, the divergencies may arise when one integrate
over $\am$. For example the UV index of the diagram (Fig.~1) is equal to
$\omega=2+2n-4deg {\cal F}_{n}\geq 2$ (here $\qm$ are the fields which
represent $\dqm$).

\begin{figure}
\caption{}
\GRAPH(hsize=4){ \mov(0.7,0.7){\dashdotlin(4.7,0)
\mov(2.3,0){\halfwavecirc (3.6)[+]}}
\ind(7,5){q}
\ind(15,5){q}
\ind(42,5){q} \ind(50,5){q} \ind(6,10){A} \ind(46,10){A}}
\end{figure}

For finite $M_{\jmath}$ this diagram is divergent. It was pointed out in
ref. \cite{SlF} that this problem will not appear if the propagators of the
fields $\am$ decrease for large momenta faster than the propagators of the fields
$\qm$. It can be achieved if the degree of covariant derivatives
in the action is higher than in the PV determinants. (Another possibility
was discussed in ref. \cite{AsF2}.)

In the next section we shall give answers to both these questions. Firstly
we demonstrate that by choosing a special form of a higher derivative
term one can avoid the problem of overlapping divergencies completely.
Secondly we present an unambigous expression for regularized functional
which does not require any additional preregularization apart from the
usual PV prescription discussed above.

\section{ Unambigous definition of regularized functional.\newline
Removing of overlapping divergencies.}

Let us choose the regularized action in the following form:

\be S^{\Lambda}=S_{YM}+\frac{1}{4\Lambda^{10}}\int [\frac{\delta S_0}
{\delta {\cal A}_{\rho}(x)}]^2 dx \label{28}\ee
where:
\be S_0=\int (\nabla^2 \fmn )^2 dx \ee
As a gauge condition we choose the regularized $\alpha$-gauge:

\be \frac{1}{4\Lambda^{10}}\int\Bigl[\nabla_{\rho} \Box^2
\pdm \am\Bigr]^2dx+\frac{1}{4\alpha}\int[\pdm \am]^2dx
\label{28a} \ee
Then the one loop functional can be written in the form:

\bea
&&\za[\am]=\int\exp \Bigl\{\frac{\imath}{2}\int\!\!\int\frac{\delta^2S_{YM}}
{\delta\ama(x)\delta{\cal A}_\nu^b(y)}\qma(x)q_\nu^b(y)dxdy+\frac{\imath}{4
\alpha}\int[\pdm \qma]^2dx+\nonumber \\&&+\frac{\imath}{8\Lambda^{10}}
\int\!\!\int\!\!\frac{\delta^2}{\delta \ama(x) \delta {\cal
A}_{\nu}^{b}(y)} \Bigl[\int\!\! dz \Bigl[\frac{\delta
S_0}{\delta {\cal A}_{\rho}(z)}+\nabla_{\rho}\Box^2\partial {\cal A}(z)
\Bigr]^2 \Bigr]\qma (x)q_{\nu}^{b}(y)dxdy
\Bigr\}\times\nonumber \\&&\times\gda\detdet \dd D \qm (x) \label{32} \eea
Note that due to the gauge invariance of $S_0$:

\be \nabla_{\rho}\frac{\delta S_0}{\delta {\cal A}_{\rho}(z)}=0 \label{sop}\ee
and the cross term in the second line of the exponent (\ref{32}) is equal to
zero. The additional determinant $\detdet$ is due to the gauge fixing
term (\ref{28a}). It is convinient for us to redefine it up to the
nonessential constant as follows:

\be \detdet\rightarrow\klio \ee

The expression (\ref{32}) is of course still formal, as it generates
ultraviolet divergent Feynman diagrams. To make the following transformations
rigorous we have to introduce some preregularization which makes
the integral (\ref{32}) meaningful. We assume that a finite lattice
a la Willson \cite{Wils} is introduced, which makes all the integrals
both ultraviolet and infrared convergent without breaking gauge invariance.
Any other gauge invariant regularization could do the job as well,
but we prefer to consider the lattice regularization as it is
universal and has a nonperturbative meaning. We emphasize that
this preregularization is needed only to make all the arguments which
follow rigorous. At the end the preregularization will be removed
and the final recipe will be formulated without any references to a
particular preregularization procedure. Having this in mind we shall not
write explicitely the corresponding lattice expressions and will keep
the continious notations assuming that:
\[\int d^4x\rightarrow a^4\sum_x\;\;\;\; ; \;\;\;\; \pdm\phi(x)
\rightarrow\frac{\phi(x+a_\mu)-\phi(x)}{a};\] e.t.c.

Differentiation of $\hact^2$ in the expression (\ref{32}) will produce two
kind of vertices:

\be\qma (x) q_{\nu}^b (y) \frac{\delta^2}{\delta \ama (x) \delta
{\cal A}_{\nu}^b (y)}\hact\times\hact \label{33} \ee
and
\be\qma(x)\frac{\delta}{\delta \ama (x)}\hact \times
\frac{\delta}{ \delta {\cal A}_{\nu}^b (y)}\hact q_{\nu}^b (y) \ee
Obviously the vertices of the type (\ref{33}) after removing a
preregularization produce only convergent diagrams as the maximal number
of derivatives acting on
the fields $\qm$ is equal to $6$, and the $q$-field propagators decrease
at $k\rightarrow \infty$ as $k^{-12}$.

Analogously one can show that the substitution:
\be \int \qm(x)\frac{\delta}{\delta \am(x)}\Bigl[
\nabla_{\rho}\Box^2\partial {\cal A}(z)\Bigr]
dx \rightarrow \nabla_{\rho}\Box^2\pdm
\qm(z) \ee affects only convergent diagrams.

Therefore being interested only in the ultraviolet divergent part of $\za$
we can rewrite it in the form:

\bea &&
\za^{div}[\am]=\int \exp \Bigl\{\frac{\imath}{4\Lambda^{10}}\int dz \Bigl[
\int\soa \qm(x)dx+\nabla_{\rho}\Box^2\pdm\qm(z)\Bigr]^2
\Bigr\}\times\nonumber\\&&\times\gda\klio \dd Dq\label{36}\eea

The notation $\za^{div}$ means that when the preregularization is
removed (i.e. in the continuum limit) the difference between the
functional $\za$, defined by the eq.(\ref{32}) and $\za^{div}$ is
ultraviolet finite. At the moment we consider the model on the
lattice, so the expression (\ref{36}) is both ultraviolet and infrared
finite by itself.

Integration of eq.(\ref{36}) over $\qm$ produces:

\bea &&det^{-1/2}\opdet^2=\nonumber\\&&=det^{-1}\opdet\eea
Hence we can rewrite the eq.(\ref{36}) in the form:

\be \za^{div}[\am]=det^{-1}\tilde{{\cal K}}\gda\klio \label{38}\ee
where:

\bea &&det^{-1}\tilde{{\cal K}}=\int\!\exp\Bigl\{\frac{\imath}{2\Lambda^5}
\int\!\! dxdz \overline{\qm}(x)\Bigl[\soa+\nm\Box^2\partial_\rho\delta(x-z)
\Bigr] q_{\rho}(y)\Bigr\}\dd D\overline{q}Dq\nonumber\\&& \label{38a}\eea
and $\overline{\qm}, \qm$ are nonhermitean commuting fields.

The eq.(\ref{38}) has a very important property. When integrated over
$\am$ it does not produce in the limit $a\rightarrow 0$ any new divergent
diagrams.  Indeed, the maximal number of derivatives acting on the fields
$\am$ is equal to $5$, and the $\am$-propagator decreases at $k\rightarrow
\infty$ as $k^{-12}$.  Therefore any diagram which contains at least one
internal $\am$-line is convergent. This is in contrast with the
expression (\ref{20}) which, as was discussed in the previous section,
being integrated over $\am$ do produce ultraviolet divergencies.
This observation solves the problem of overlapping divergencies in the higher
covariant derivative regularization.

Now we are ready to derive an unambigous expression for the regularized
functional. Let us write the functional generated by the regularized action
(\ref{28}) and the gauge fixing term (\ref{28a}) in the form:

\bea &&\za[J]=\int\exp\Bigl\{\imath S_{YM}+\imath\int\Bigl[\frac{1}{4\alpha}
(\pdm\am)^2+J_{\mu}\am-\frac{1}{2}h_{\rho}^2\Bigr]dx+\nonumber\\&&+
\frac{\imath}{2\Lambda^5}\int h_{\rho}(z) \hact dz\Bigr\}\Bigl[det {\cal K}
\cdot det^{-1}{\cal K}\Bigr]\times\nonumber\\&&\times\gda\klio\dd Dh_\rho
D\am\label{40a}\eea
Indeed, integrating over auxiliary fields $h_{\rho}$ and taking into
account the identity (\ref{sop}) we get the functional corresponding to
the regularized action (\ref{28}) with the gauge fixing term (\ref{28a}).
Here $det {\cal K}$ is defined by the equation similar to eq.(\ref{38a}):

\bea &&det^{-1}{\cal K}=\int\exp\Bigl\{\frac{\imath}{2}\int\!\!\int
dxdz\overline{\qm}(x)\Bigl[\frac{1}{\Lambda^5}\soa+\frac{1}{\Lambda^5}
\nm\Box^2\partial_\rho\delta(x-z)
+\nonumber\\&&+\underbrace{\cov+\frac{\nm}{\Lambda}\partial_{\rho}\delta(x-z)}
\Bigr]q_\rho(z)\Bigr\}\dd D\overline{q}Dq \label{38b}\eea
This expression differs from (\ref{38a}) by underbraced terms which are
introduced to provide the infrared convergence of $det^{-1}{\cal K}$
in the limit when the preregularization is removed. They do not
influence the ultraviolet asymptotics, and the ultraviolet divergent
parts of $det {\cal K}$ and $det \tilde{{\cal K}}$ coincide.

Let us show that in the limit $a\rightarrow 0$ $det {\cal K}$ exactly
compensates the one loop divergencies generated by the integration
of the exponeent in eq.(\ref{40a}).

The free propagators generated by the exponent in the eq.(\ref{40a})
have the following UV behaviour: $\widehat{\am {\cal A}_\nu}\sim k^{-12}$;
$\widehat{h_{\rho}h_\sigma}\sim k^{-10}$; $\widehat{\am h_\rho}\sim k^{-6}$.
One sees that as soon as a diagram includes at least one $\widehat{\am
{\cal A}_\nu}$ or $\widehat{h_\rho h_\sigma}$ propagator it is convergent.
The only divergent diagrams are one loop diagrams formed by the propagator
$\widehat{h_{\rho}\am}$. The sum of these diagrams can be presented as
follows:

\be\int\exp\Bigl\{\frac{\imath}{2\Lambda^5}\int\!\!\int dxdzh_\rho(z)
\frac{\delta}{\delta\am(x)}\hact \qm(x)\Bigr\}\dd DhDq\label{43b}\ee
and according to the discussion given above coincides up to the finite
terms with $det^{-1} \tilde{{\cal K}}$. The divergent diagrams generated by
eq.s.(\ref{38b}) and (\ref{43b}) have the same structure and therefore to
provide the existence of the limit $det^{-1}\tilde{{\cal K}} det{\cal K}$
when the preregularization is removed it is sufficient to impose the usual
PV prescription.

The only divergent factors in the eq.(\ref{40a}) are:

\be det^{-1}{\cal K} \gda\klio \label{44b}\ee
Let us show that in analogy with the heuristic arguments of the preceeding
section this expression can be written in the form which makes it's
divergent part manifestly gauge invariant.

The expression (\ref{44b}) can be rewritten in the form:

\bea &&\za^{div}[\am]=
\int\exp\Bigl\{\frac{\imath}{2}\Bigl[\frac{1}{\Lambda^5}\int\!\!\int
\overline{\qm}(x)\soa q_{\rho}(z)dxdz+
\nonumber \\&&+\frac{1}{\Lambda}\int\Bigl[-\nm\overline{\qm}^a W^a
+\overline{q_\rho}(\nabla_\rho \nm-\nabla^2
g_{\rho\mu})\qm\Bigr] dx\Bigr]\Bigr\}\gda\times\nonumber \\&&\times
\klio \dd \delta(\Bigl(\frac{\Box^2}{\Lambda^4}+1\Bigr)\pdm\qm-W)D
\overline{q} Dq DW\eea
To pass to a covariant background gauge we multiply $\za^{div}$ by "unity":
\be det \nabla^2 det\finv\int\dd \delta (\finv\nm(\qm+\nm u)-W(x))Du(x)=1
\ee
(where $\finv=\frac{\nabla^4}{\Lambda^4}+1$) and make the change of
variables:  \be \qm\rightarrow \qm-\nm u \ee After these transformations
one can integrate over $u, W$ to get:

\bea
&&\za^{div}=\int\exp\Bigl\{\frac{\imath}{2\Lambda^5}\int\!\!\int
\overline{\qm} (x)\soa
(q_{\rho}-\nabla_{\rho}u)_zdxdz+\nonumber\\&&+\frac{\imath}{2\Lambda}\int
\Bigl[-\nm\overline{\qm}\finv\nabla_{\rho}q_{\rho}
+\overline{\qm}(\nm\nabla_\rho
-\nabla^2g_{\mu\rho})q_\rho\Bigr] dx\Bigr\}\times\nonumber\\&&\times
det\nabla^{2}\det\finv \klio\dd D\overline{q}Dq \label{44}\eea
where:
\[u={\cal M}^{-1}\Bigl[\pdm\qm-\Bigl(\frac{\Box^2}{\Lambda^4}+1
\Bigr)^{-1}\finv\nm\qm\Bigr]\]
In the same way as it has been done in the section 2 one can show that the
terms proportional to $\nabla_{\rho}u$ in the limit when the lattice
preregularization is removed $(a\rightarrow 0)$ generate only convergent
diagrams. Therefore apart from the factor $\klio$ the divergent in the
limit $a\rightarrow 0$ part of $\za^{div}$ has a manifestly gauge invariant
form.

Although $\klio$ is not manifestly gauge invariant, it can be
regularized in a gauge invariant way. To make this determinant finite in the
limit $a\rightarrow 0$, it is sufficient to multiply it by gauge invariant
PV determinants:

\be\klio\rightarrow\klio\prod_{\jmath}det^{\cj/2}(\nabla^2-M_\jmath^2) \ee

Therefore to get the functional which is finite in the limit $a\rightarrow
0$ we can introduce a gauge invariant interaction of PV fields:

\bea &&I_{PV}=\int\exp\Bigl\{\frac{\imath}{2}\int\!\!\int
\overline{B}_{\mu}(x)\Bigl[\indet\Bigr]B_{\rho}(z)dxdz+\nonumber \\ &&
+\imath M\int\overline{B}_{\mu}^aB_{\mu}^a dx\Bigr\}\dd D\overline{B} DB
\prod_{\jmath}det^{3/2\cj}(\nabla^2-M_{\jmath}^2)det^{-1}\finv \label{46}\eea
Here $\overline{B}_{\mu}, B_{\mu}$ are anticommuting PV fields, and
conditions (\ref{21},\ref{21a}) hold.

Integration over $\overline{B}_{\mu}, B_{\mu}$ will produce the factor:

\bea &&I_{PV}=det \Bigl[\indet +Mg_{\mu\rho}\delta(x-z)\Bigr]\times
\nonumber\\&&\times\prod_{\jmath} det^{3/2\cj}(\nabla^2-M_{\jmath}^2)
det^{-1}\finv \label{52b}\eea
which compensates the divergencies of the functional (\ref{44}).
Obviously in the limit \newline $M\rightarrow\infty, \Lambda\rightarrow\infty$
all unphysical exitations decouple.

One sees that the sum of the diagrams generated by the functionals
(\ref{52b}), (\ref{44}) has a finite limit when $a\rightarrow 0$. Moreover in
the limit $a\rightarrow 0$ the divergent diagrams generated by the functionals
(\ref{52b}) and (\ref{44}) have the same structure. So the auxiliary
lattice preregularization can be omitted and to make the
sum finite it is sufficient to use the standard PV prescription: the momenta
in the similar diagrams have to be assigned in the same way.
It allows to write an unambigous finite expression for the regularized
functional. As was shown above for a finite lattice preregularization
we can replace in the functional (\ref{40a}) the factor:

\be det^{-1}{\cal K}\gda\klio \ee
by the expression (\ref{44}), regularized by adding the gauge invariant
PV terms (\ref{52b}). After that we can remove a preregularization.
The limiting expression is finite provided the usual PV prescription
is used. In other words one can forget about a preregularization at all
and take the expression obtained in this way as a definition of the regularized
functional.

Before writing the final result let us note that the nonlocal term in the
eq.(\ref{44}) proportional to ${\cal M}^{-1}$ does not contribute in the
limit $\Lambda\rightarrow\infty$. For finite $\Lambda$ we have shown
it produces finite diagrams, and in the limit $\Lambda\rightarrow\infty$
its contribution disappears. Being interested finally in the limit
$\Lambda\rightarrow\infty$ we can omit this term in the eq.(\ref{44}).
It simplifies the expression for the regularized functional and make
the effective action local. Omitting this term we break the gauge
invariance for finite $\Lambda$. The Slavnov-Taylor identities will be
violated by finite terms of order $O(\Lambda^{-1})$. These terms are harmless
as they have no influence on the counterterms and disappear in the limit
$\Lambda\rightarrow\infty$.

Having in mind these remarks we can write the unambigous expression for
the regularized functional which does not require any special
preregularization procedure. It looks as follows:

\bea &&{\cal Z}_{\Lambda,M}[J]=\int\exp\Bigl\{\imath S_{YM}+\frac{\imath}
{2\Lambda^5}\int\hact h_{\rho}(z)dz+\nonumber\\&&
+\int[\frac{1}{4\alpha}(\pdm\am)^2+J_{\mu}{\am}-\frac{1}{2} h_{\rho}^2(x)]dx
+\frac{\imath}{2}\int\!\!\int\overline{b}_{\mu}(x)\Bigl[\frac{1}{\Lambda^5}\soa
+\nonumber\\&&+\frac{\nm}{\Lambda}\Bigl(\frac{\Box^2}{\Lambda^4}+1\Bigr)
\partial_\rho\delta(x-z)+\frac{1}{\Lambda}(\nm\nabla_\rho-\nabla^2g_{\rho
\mu})\delta(x-z)\Bigr]b_{\rho}(z)dxdz\Bigr\}\times\nonumber\\&&\times
\exp\Bigl\{\frac{\imath}{2}\int\!\!\int\overline{\qm} (x)\Bigl[
\frac{1}{\Lambda^5}\soa+\frac{\nm}{\Lambda}\finv\nabla_\rho\delta(x-z)
+\nonumber\\&&+\frac{1}{\Lambda}(\nm\nabla_\rho-\nabla^2g_{\mu\rho})\delta(x-z)
\Bigr]q_{\rho}(z)dxdz+\frac{\imath}{2\Lambda}\int\!\!\int\overline{B}_{\mu}(x)
\Bigl[\frac{1}{\Lambda^4}\soa+\nonumber \\&&+\nm\finv\nabla_{\rho}
\delta(x-z)\Bigr]B_{\rho}(z)dxdz+\imath M\int\overline{B}_{\mu}B_{\mu}dx\Bigr\}
\klio \times\nonumber\\&&\times det\nabla^2\prod_{\jmath}
det^{3/2\cj}(\nabla^2-M_{\jmath}^2) \dd Dh_{\rho}D\overline{b}_{\mu}Db_{\mu}
D\overline{\qm}D\qm D\overline{B}_{\mu}DB_{\mu}D\am \label{48}\eea
This rather lengthy expression has in fact a simple meaning. The integral
over the anticommuting fields $\overline{b_\mu}, b_\rho$ subtract the
divergent one loop diagrams which arise due to integration over $\am,
h_\rho$. The integral over PV fields $\overline{B_{\mu}},B_\rho$ subtract
analogous divergencies which arise due to integration over fields
$\overline{\qm},q_\rho$. As the propagators of the fields $\am$
decrease for $k\rightarrow\infty$ as $k^{-12}$, no overlapping divergencies
are present.

Let us remind how the expression (\ref{48}) was obtained. We firstly
transformed identically the preregularized functional which satisfied
the correct Slavnov-Taylor identities. Then we multiplied it by a gauge
invariant factor depending on PV fields. Obviously the resulting
functional satisfies the same identities for any finite preregulator $a$,
and as the limit $a\rightarrow 0$ exists, in the limit $a\rightarrow 0$ as
well. The only procedure which could break the gauge invariance was
omitting of the nonlocal term $\sim{\cal M}^{-1}$. But as we discussed
above it has no influence on the final result.

The functional (\ref{48}) has unambigous meaning as all the diagrams
generated by the expansion of eq.(\ref{48}) are finite, provided the
momenta of similar diagrams are assigned in the same way.
Therefore as we have already discussed we can take it as a definition of
regularized functional forgetting completely about a preregularization.

\section{Discussion.}

In this paper we showed that contrary to the statement of the authors
\cite{Ruiz} the higher covariant derivative regularization supplemented
by the additional PV type regularization of one loop diagrams do provide
a consistent regularization of QCD and other gauge invariant models.
It can be used as a practical method of calculations in theories where
dimensional regularization is not applicable, and may also serve as a
starting point for nonperturbative approaches. The formulation of the method
given in the present paper avoids the problem of overlapping divergencies
and gives unambigous method of calculations which do not require any
additional preregularization.

  \end{document}